\documentclass[aps,prl,twocolumn,showpacs]{revtex4}
\usepackage{amsmath,amssymb,epsf,bm}
\usepackage{graphicx}
\begin{document}
\title{Quantum oscillations of spin current through a III-V semiconductor
loop}
\author{A. G. Mal'shukov}
\affiliation{Institute of Spectroscopy, Russian Academy of Science,
142190 Troitsk, Moscow oblast, Russia}
\author{V. Shlyapin}
\affiliation{Institute of Spectroscopy, Russian Academy of Science,
142190 Troitsk, Moscow oblast, Russia}
\author{K. A. Chao}
\affiliation{Solid State Theory, Department of Physics, Lund University,
S-223 62 Lund, Sweden}

\begin{abstract}
We have investigated the transport of spin polarization through a
classically chaotic semiconductor loop with a strong Rashba spin-orbit
interaction. We found that if the escape time of a particle is long enough,
the configuration averaged spin conductance oscillates strongly with the
geometric spin phase. We predict a sizable rotation of spin polarization
along its flowing path across the loop from the injector to the collector.
We have also discovered a quantized universal spin relaxation in a 2D
reservoir connected to such a semiconductor loop.
\end{abstract}
\maketitle

In the emerging field of spin electronics, the recent achievement of spin
injection into paramagnetic semiconductors~\cite{inject} makes it an urgent
task to control the spin current in semiconductor nanostructures. The spin
current in a 2D channel of narrow gap III-V semiconductor can be manipulated
by taking advantage of the strong spin-orbit splitting of the conduction
electron energy, because the mechanism of such splitting produces a spin
precession which depends on the electron quasimomentum. One example is the
spin valve transistor~\cite{datta}, in which spin polarisation precesses in
a 2D semiconductor channel between a ferromagnet spin injector and a
ferromagnetic spin collector. The measured resistance is determined by the
angle of spin rotation along the propagating path. This angle can be varied
by adjusting the spin-orbit interaction (SOI) strength in the semiconductor
with an external gate~\cite{nitta}.

In this Letter we will investigate an interesting phenomenon which can be
observed in a 2D semiconductor loop as shown in Fig.~1. It is well
known~\cite{ms} that due to the SOI, when an electron travels along a
closed path, its wave function accumulates an additional phase $\psi$. If
the SOI is a linear function of the electron quasimomentum, this phase
depends only on the shape and the length of the path. In multiconnected
conductors the effect of this phase on electron transport is similar to the
Aharonov-Bohm (AB) effect. For example, in a disordered 2D ring, $\psi$
adds itself to the AB phase in the Aronov-Altshuller-Spivak oscillation of
the DC electric conductance~\cite{alt}, as well as to the AB oscillation of
the electric conductance mesoscopic fluctuations~\cite{meso}.
\begin{figure}
\includegraphics[width=8.4 cm]{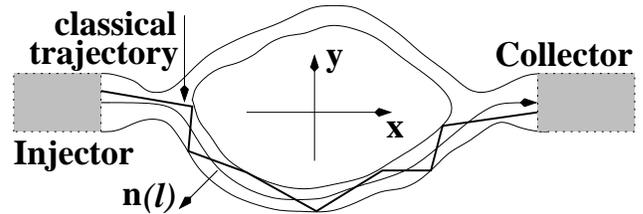}
\caption{A schematic plot of the loop sample. The zigzag line represents
classical trajectory $a$, and the arrowed curve is the smooth $l$-path.
$\bm{n}(l)$ is a vector normal to the $l$-path.}
\label{fig1}
\end{figure}

In this Letter, instead of the oscillation of the electric current, we will
study the effect of the spin phase on the quantum oscillation of the spin
current. However, we will assume the motion of electrons as ballistic along
their classical trajectories, rather than a diffusive transport inside the
loop~\cite{ms,meso}. One such trajectory is schematically illustrated by
the zigzag-lined path in Fig.~1, although it can be curved by a smooth
random potential produced by the modulation doped impurities. We assume that
the motion of a particle inside the loop is classically chaotic, and so the
quasiclassic approach of Ref.~\cite{JBS} can be applied. Nevertheless, to
calculate the spin current through the loop we need to generalize this
method by taking into account the spin degree of freedom and the SOI. With
this approach, we will calculate the average spin current. This implies that
its mesoscopic fluctuations will be averaged out. The corresponding
experimental performance is, for example, to average the results measured
with several gate voltage sweeps. We will also ignore the weak localization
correction, which is small because our system has a large number of
transport channels through the loop. We will show that the so calculated
spin conductance oscillates as a function of the SOI strength, and
consequently can be controlled by varying the gate voltage. We would like
to emphasize that these quantum oscillations appear in the classical spin
conductance, which is a drastically different phenomenon from the AB
effect. The AB effect is abcent in the average spin current when the weak
localization effects are ignored.

Following the Landauer approach, to study the spin dependent conductance
we define
\begin{equation}\label{g}
g_{\alpha\beta\gamma\delta} =
\frac{e^2}{h} \sum_{n,m} t_{nm}^{\alpha\beta}t_{nm}^{\gamma\delta\ast} \, ,
\end{equation}
where $t_{nm}^{\alpha\beta}$ is the transmission amplitude of an electron
at Fermi energy $E_F$ propagating from the channel $n$ and the spin state
$\alpha$ in the injector to the channel $m$ and the spin state $\beta$ in
the collector. $t_{nm}^{\alpha\beta}$ itself is the $\alpha\beta$--element
of the matrix $t_{mn}$ which operates on spin states. The usual spin
independent electric conductance is simply 
$g$=$(e^2/h)\sum_{n,m,\alpha,\beta}|t_{nm}^{\alpha\beta}|^2$. If a spin
orinented along the $x$-axis is injected from the injector, and its
orientation becomes along the $y$-axis when the spin is collected at the
collector, let $g_{xy}$ represent this spin current passing through the
loop. The matrix elements can then be written as
\begin{equation}\label{gij}
g_{ij}=\frac{e^{2}}{h}\sum_{n,m}Tr\left\{ \sigma _{i}t_{nm}\sigma
_{j}t_{nm}^{\dag}\right\} \, ,
\end{equation}
where $\sigma_i$ are Pauli matrices with $i$=$x,y,z$. The spin orientation
can be detected by measuring the polarization of the emitted
light~\cite{inject}. In such an experiment, the polarization matrix of the
emitted photons can be derived if we know $g_{ij}$.

In narrow gap III-V semiconductor quantum wells, the SOI is dominated by
the Rashba interaction~\cite{rash} with the interaction Hamiltonian
\begin{equation}\label{hso}
H_{so} = \alpha \, \bm{\sigma} \times {\bf p} \, ,
\end{equation}
where {\bf p} is the momentum operator, and the vector $\bm{\sigma}$ has
components $\sigma_x$, $\sigma_y$, and $\sigma_z$. Then the quasiclassical
expression of $g_{\alpha\beta\gamma\delta}$ can be easily obtained as
\begin{equation}\label{g2}
g_{\alpha\beta\gamma\delta} =
\frac{e^2}{h}\sum_{a,b} t_0(a)t_0^{\ast}(b)
S_a^{\alpha\beta}S_b^{\gamma\delta\ast} \, ,
\end{equation}
where $t_0(s)$ is the spin independent transmission amplitude for a
classical trajectory labelled by $s$. One such trajectory is schematically
plotteded in Fig.~1 as the zigzag line. The explicit expression of $t_0(s)$
as well as the boundary conditions are given in Ref.~\onlinecite{JBS}. The
spin evolution operator $S_a$ along the $a$--trajectory is defined as
\begin{equation}\label{s}
S_a = T \left[ \exp \left\{
-\frac{i}{\hbar} \alpha \, m^{\ast} \int_a {\bf z} \times
\bm{\sigma}({\bf r}) \, d{\bf r} \right\} \right] \, ,
\end{equation}
where the integration is along the $a$--trajectory, $m^{\ast}$ is the
effective mass, and {\bf z} is a unit vector parallel to the z-axis. The
symbol $T$ means that the Pauli matrices must be ordered along the path.

The $g_{\alpha\beta\gamma\delta}$ given in (\ref{g2}) must be averaged over
the mesoscopic fluctuations, using the procedure described in
Ref.~\onlinecite{JBS}. Such averaging may be considered as a temperature
effect, or, in accordance with the ergodic hypothesis, as an average over
an ensemble of loops of slightly different shapes. Because of the rapidly
oscillating phase factors in the quasiclassical amplitudes $t_0$, after the
averaging procedure, in (\ref{g2}) only the terms with $a$=$b$ remain.
Accordingly, we obtain from (\ref{gij}) the so averaged spin conductance 
$\left< g_{ij}\right>$ as
\begin{equation}\label{gij2}
\left<g_{ij}\right> = \frac{e}{h}\sum_a|t_0(a)|^2 D_{ij}^a\ .
\end{equation}  
where 
\begin{equation}\label{Dij}
D_{ij}^a = Tr\left\{\sigma_i S_a\sigma_j S_a^{\dag} \right\}\ .
\end{equation}
Similarly, the so averaged electrical conductance is simply
$(2e/h)\sum_a|t_0(a)|^2$, and is spin independent.

The evolution matrix can be parametrized using its property that it is a
SU(2) representation of 3D rotations. In fact, we can express $S_{a}$ as
a time ordered product of infinitesimal rotations corresponding to small
shifts $d${\bf r} along the trajectory $a$. Each infinitesimal rotation is
along the axis $d${\bf r}$\times${\bf z} through an angle
2$|d${\bf r}$\times${\bf z}$|/L_{so}$. These infinitesimal rotations are
represented by operators
$\exp [(-i/L_{so})(d{\bf r}\times {\bf z})\bm{\sigma}]$, and they sum up
to make $S_{a}$ for a finite rotation through the angle $2\psi_a$ around a
unit vector ${\bf N}_a$. Hence, the evolution matrix in (\ref{s}) can be
represented as 
\begin{equation}
S_a=e^{i \psi_a{\bf N}_a\bm{\sigma}}\ .
\label{s2}
\end{equation}
We should notice that $\psi_a$ and {\bf N}$_a$ are uniquely determined by
the geometric shape and the length of the trajectory $a$.

From now on we will consider a particular sample geometry that the area
occupied by the 2D electron gas in the loop is much less than $L_{so}^2$,
and the linear dimension of the loop can be larger than $L_{so}$, where
$L_{so}$=$\hbar/\alpha m^*$. In other word, both the upper path and the
lower path of the loop are narrow. In this case one can show that each
trajectory $a$ in (\ref{s}), as indicated by the zigzag line in Fig.~1, can
be replaced by a smooth trajectory, which is shown in Fig.~1 as the arrowed
curve. We will label this smooth curve as $l$-path. Let $\theta_a$ be the
area enclosed by the classical trajectory $a$ making one turn around the
loop. Then, the area enclosed by the $l$-path is the average of $\theta_a$
over classical trajectories. Deviations of real paths from the $l$-path can
be treated perturbatively, which will be reported elsewhere.

After the trajectory $a$ is replaced by the $l$-path, the evolution matrix
(\ref{s}) becomes a simple function of the number $w$ of windings the
trajectory $a$ makes around the loop until a particle escapes into the
collector. $w$ is positive if the winding is counterclockwise. The
corresponding evolution operator for the smooth $l$-path, denoted as
$S(w)$, can be expressed as 
\begin{equation}\label{s3}
S(w) = e^{i \psi_0 {\bf N}_0 \bm{\sigma}}
e^{iw \psi {\bf N} \bm{\sigma}} \, .
\end{equation} 
Here $\psi_0$ and {\bf N}$_0$ are the 3D rotation parameters for the
$l$-path in the lower half of the loop, and $\psi$ and {\bf N} are the 3D
rotation parameters for the $l$-path around the complete loop. Using
Eq.~(\ref{s3}) we obtain the general dependence of the trace in (\ref{Dij})
on the winding number
\begin{equation}
D_{ij}(w) = M_{ij}^{(1)} e^{i2w\psi} + M_{ij}^{(-1)} e^{-i2w\psi} +
M_{ij}^{(0)} \, , 
\label{Dij2}
\end{equation} 
where the matrix elements $M_{ij}^{(1)}=M_{ij}^{(-1)*}$ depend only on the 
geometric shape of the $l$-path.

Based on the above expressions, we can follow the approach used in
Ref.~\onlinecite{Kaw} to calculate the Aharonov-Bohm effect on mesoscopic
electric conductance fluctuations in doubly connected classically chaotic
loop. Let $T$ be the time interval that a particle spends inside the loop,
and $T_0$ be such a duration for the shortest trajectory. Then, according
to Ref.~\onlinecite{Kaw}, we average the winding number with the Gaussian
distribution function
\begin{equation}\label{Gauss}
{\cal P}(w|T) = \sqrt{\frac{T_0}{2\pi\alpha T}}
\exp \left\{ -\frac{w^2}{2\beta T/T_0} \right\} \, ,
\end{equation} 
where $\beta$ is the system dependent dimensionless constant. 
For classically chaotic systems, it has been shown~\cite{JBS} that $T$
obeys the distribution function
${\cal P}(T)$=$\tau^{-1}\exp [-(T-T_0)/\tau]$, where $\tau$ is the mean
escape time of the particles. We should point out that this kind of
statistical approach is valid only for large $\tau$, such that the particle
can travel around the loop many times before it escapes from the loop.
Hence, we assume $\kappa$$\equiv$$\sqrt{2T_0/\tau\beta}$$\ll$1. After
averaging (\ref{Dij2}) over $w$ and $T$ we arrive at 
\begin{eqnarray}\label{Dij3}
&& \left< D_{ij} \right>_{w,T}  = M_{ij}^{(0)} +
2\,Re[M_{ij}^{(1)}]\,A(\psi) \\ 
&& A(\psi) = \kappa^2/(\kappa^2+4\sin^2 \psi) \, . \nonumber
\end{eqnarray}
The oscillation pattern of the spin current, caused by the $A(\psi)$ term,
gets sharper as $\kappa$ becomes smaller, and eventually transforms into a
periodic array of narrow peaks at positions $\psi$=$\pi n$.

Eq.~(\ref{Dij3}) provides a general relation between the spin conductance and
the spin phase $\psi$. Here we will consider a
specific example that the loop is nearly circular of radius $d$, and is
symmetrically attached to two leads. For this geometry, by taking proper
derivative with respect to the length of the $l$-path, we obtain from
(\ref{s}) the differential equation
\begin{equation}\label{difeq}
\frac{dS}{dl} = -\frac{i}{L_{so}}({\bf n}(l) \cdot \bm{\sigma}) S \, ,
\end{equation}
where the vector ${\bf n}(l)$ is shown in Fig.~1. For a cicular $l$-path,
this equation is equivalent to the time-dependent Schr\"odinger equation
for a $\frac{1}{2}$ spin interacting with a uniformly rotating planar
magnetic field. The solution of this equation is well known~\cite{Landau}.
Let us define $\gamma$=$d/L_{so}$, and $g_0$ the spin independent
conductance of the loop. Referring to the coordinate system specified in
Fig.~1, we obtain $\psi$=$\pi\sqrt{1+4\gamma^2}$, and the nonzero
components of the spin conductance as
\begin{eqnarray}\label{fing}
\left< g_{xx} \right>
&=& -g_0(\pi^2/\psi^2)\,[4\gamma^2 + A(\psi)\cos\psi] \nonumber \\
\left< g_{xz} \right> &=& -\left< g_{zx} \right> =
g_0(2\gamma\pi^2/\psi^2)\,[1- A(\psi)\cos\psi] \nonumber \\
\left< g_{yy} \right> &=& -g_0 \,A(\psi)\cos\psi \nonumber \\
\left< g_{zz} \right>
&=& g_0(\pi^2/\psi^2)\,[1+4\gamma^2 A(\psi)\cos\psi] \, .
\end{eqnarray}
From these expressions it is obvious that the oscillations of $g$ can be
observed as long as $d$$\geq$$L_{so}$. From Ref.~\onlinecite{nitta} we
evaluate $L_{so}$$\simeq$3000~\AA$\,$ in InGaAs/InAlAs quantum wells.
Hence, the loop size must be larger than about 1~$\mu$m but less that the
electron dephasing length, which can be very long at low temperatures. 

So far we have considered the spin transport between the injector and the
collector through the loop. The mechanism of transport process involves
spin diffusion driven by the difference of spin polarizations between the
lead attached to the injector and the lead attached to the collector. Such
a transport is represented by the spin conductance (\ref{g}) which, is
determined by the spin dependent transmittance of the loop. Besides this
process the SOI in the loop gives rise to a spin dependent reflectance. A
particle which enters the loop from the injector-connected lead with a
given spin orientation can be reflected back into the same lead with an
opposite spin direction. This provides an additional relaxation process of
the spin polarization in this lead, and the oscillatory dependence of such
a relaxation on the spin phase $\psi$ is expected. A very suitable system
for studying this relaxation mechanism is a reservoir connected to a loop
via a point contact which has ${\cal N}$ transmitting channels. The
reservoir needs not to be very big. At a certain time, a nonequlibrium
spin polarization  $\Sigma$=($N_{\uparrow}$--$N_{\downarrow}$)/2 is
created in the reservoir, where $N_{\sigma}$ is the number of particles
with spin projection $\sigma$ onto the quantization axis. Let
$R_{\uparrow\downarrow}$ be the reflectance associated to the electron spin
flip reflection. Then, the time rate of change of $\Sigma$ is given by
\begin{equation}\label{P}
\frac{d\Sigma}{dt} =
-\frac{1}{2h}R_{\uparrow\downarrow}\,(\mu_{\uparrow}-\mu_{\downarrow})
-\frac{\Sigma}{\tau_{s}} \, ,
\end{equation} 
where $\mu_{\sigma}$ is the chemical potential of the $\sigma$-spin state,
and $\tau_{s}$ is the spin relaxation time within the reservoir.

Let the reservoir be a 2D degenerate electron gas of volume $V$, with the
density of states at the Fermi level $N(E_F)$=$m^*/2\pi\hbar^2$. Then,
$(\mu_{\uparrow}$--$\mu_{\downarrow})$=2$\Sigma$/$N(E_F)V$, and we find
from (\ref{P}) an additional spin relaxation rate in the reservoir
\begin{equation}\label{gamma}
\Gamma =(\hbar/m^*V)\,R_{\uparrow\downarrow} \, ,
\end{equation}  
due to its connection to the loop. Followinging the same quasiclassical
approach which leads to the results (\ref{fing}), we obtain
\begin{equation}\label{R}
R_{\uparrow\downarrow} = {\cal N}
\left< |\left( e^{\,iw\psi{\bf N} \bm{\sigma}}
\right)_{\uparrow\downarrow}|^2 \right>_w \, ,
\end{equation}
where $<\cdots >_w$ is an average over the winding number $w$. If the
direction of spin polarization is parallel to {\bf N}, then, the evolution
matrix in (\ref{R}) is diagonal, and hence $\Gamma$=0. On the other hand, 
if the spin polarization is perpendicular to ${\bf N}$, we get
$|(\exp[iw\psi{\bf N}\bm{\sigma}])_{\uparrow\downarrow}|^2$=$\sin^2 w\psi$.
Consequently, averaging over $w$ gives the final form for the relaxation
rate
\begin{equation}\label{gamma2}
\Gamma = ({\cal N}\hbar/m^*V)\,[1-A(\psi)] \, .
\end{equation}

For small $\kappa$ the function [1--$A(\psi)$] oscillates between zero for
$\psi$=$n\pi$ and a value very close to 1 for
$\psi$=($n$+$\frac{1}{2}$)$\pi$. Hence, around
$\psi$=($n$+$\frac{1}{2}$)$\pi$ the spin relaxation rate $\Gamma$ is
determined mainly by the ratio ${\cal N}/V$. Taking a typical value
$V$=4~$\mu$m$^2$, and the electron effective mass $m^*/m$=0.03 as in InAs,
we obtain $\Gamma$$\simeq$${\cal N}\cdot$10$^9$~sec$^{-1}$. The "quantum"
of relaxation time, 10$^{-9}$~sec, is comparable to the intrinsic
relaxation time $\tau_{s}$. We would like to emphasize that at this fixed
spin phase $\psi$=($n$+$\frac{1}{2}$)$\pi$, the relaxation rate is a
universal value in the sense that it does not depend on a shape and other
parameters of the loop. On the other hand, we should also point out that
(\ref{gamma2}) is valid only in the quasiclassic limit when
${\cal N}$$\gg$1. Therefore, one can not extrapolate this result to
${\cal N}$$\simeq$1 where this quantization effect may be observed
experimentally.

In order to be able to observe the spin relaxation due to the SOI in the
loop, it is necessary to have a long intrinsic spin relaxation time in the
reservoir. Therefore, the SOI in the reservoir must be sufficiently small
while the SOI in the loop must be sufficiently large. Such a condition can
be created by proper modulation doping, resulting in a highly asymmetric
confinement potential near the loop but a nearly symmetric confinement
potential in the vicinity of reservoir. A strong Rashba SOI will then
appear only in the loop.
 
The work was supported by the Royal Swedish Academy of Sciences and the
Crafoord Foundation.

\end{document}